# Geographies of an online social network

**Balázs Lengyel[a,b, 1], Attila Varga[c], Bence Ságvári[a,d,] Ákos Jakobi[e], János Kertész[f,g]**

[a] International Business School Budapest, OSON Research Lab, Záhony utca 7, 1031 Budapest, Hungary
[b] Hungarian Academy of Sciences, Centre for Economic and Regional Studies, Budaörsi út 45, 1112 Budapest, Hungary
[c] The University of Arizona, School of Sociology, Tucson, AZ 85721-0027, USA
[d] Hungarian Academy of Sciences, Centre for Social Sciences, Országház utca 30, 1014 Budapest, Hungary
[e] Eötvös Loránd University, Department of Regional Science; Pázmány Péter sétány 1/C, 1117 Budapest, Hungary
[f] Central European University, Centre for Network Science, Nádor utca 9, 1051 Budapest, Hungary
[g] Budapest University of Technology and Economics, Institute of Physics, Budafoki út 8, 1111 Budapest, Hungary

[1] Corresponding author: blengyel@ibs-b.hu.

# Geographies of an online social network

## Abstract

How is online social media activity structured in the geographical space? Recent studies have shown that in spite of earlier visions about the "death of distance", physical proximity is still a major factor in social tie formation and maintenance in virtual social networks. Yet, it is unclear, what are the characteristics of the distance dependence in online social networks. In order to explore this issue the complete network of the former major Hungarian online social network is analyzed. We find that the distance dependence is weaker for the online social network ties than what was found earlier for phone communication networks. For a further analysis we introduced a coarser granularity: We identified the settlements with the nodes of a network and assigned two kinds of weights to the links between them. When the weights are proportional to the number of contacts we observed weakly formed, but spatially based modules resemble to the borders of macro-regions, the highest level of regional administration in the country. If the weights are defined relative to an uncorrelated null model, the next level of administrative regions, counties are reflected.

# 1. Introduction

Tie formation and maintenance requires investment from both parties, and physical distance is one of the main limitations of sustaining relationships. Ties between people are less likely to be established or they need more effort to be built up as distance grows [1, 2]. The most effective way to keep up with connections is face to face communication which is greatly hindered by distance [3]. However the effect of distance and geographical regions is decreasing with the advent of printing, and telecommunication in the modern era. Modern Information Communication Technology (ICT) influences our social behavior like tie formation; consequently, it has to have a major effect on the structure of social networks [4]. This influence is clearly present in the vanishing distance dependent costs of online telecommunication, leading to the claim of the "*Death of Distance*" thesis [5] that is geographical distance less of a constraining factor on interpersonal relationships and business life in the new era of ICT.

However, some studies point to another, perhaps less intuitive direction. Goldenberg and Levy [6] argued that the ease of communication, due to new technologies, enhances the already existing ties more leading to even stronger distance dependence for online contacts then observed in the offline world. In a large, comparative study Mok et al [7] examined the email, phone, face-to-face and overall contacts before and after the internet revolution and concluded that "the sensitivity of these relationships to distance has remained similar, despite the communication affordances of the Internet and low-cost telephony". However, scholars also warn us that the role of geography as the primer dimension of community formation might be overestimated [8]. These "maintenance costs" may constitute a major hindering factor even in the case of internet-based communication [9], but its' relatedness to geography needs further investigation.

The purpose of the present paper is to contribute to this ongoing debate by assessing the effect of geographical factors on online social networking structures at the national level in Hungary. Online social networks (OSN) are products of internet-based communication, when people document and maintain friendship with those they have known each other primarily in real life [10]. OSNs have been found similar to offline social networks in terms of degree distribution [11-13]. However, there is a crucial difference between these networks: users are not able to maintain strong relations with all of those they document friendship with in an OSN environment. Evidence shows that the Social Brain Hypothesis applies in the internet era too and the size of ego-networks are limited by the capacity of human brain [14-16] as well as other limitations like available time [17]. Only few of hundreds and thousands documented online friends might be maintained. To put it differently, strong and weak ties are usually mixed in OSNs, which means less maintenance costs per documented friends than in other networks. Thus, one may expect smaller distance-decay and looser spatial dependence than reported earlier on telephone-call networks [18-19] unless other information than the list of friends is available.

Previous research has found that the probability of OSN friendship decreases as distance grows [20, 21], which makes the majority of links concentrate in closed geographical areas [22,13]. Establishing an OSN tie is also related to other types of costs such as breaking through cultural and language barriers [23], thus geographical properties remain very important when navigating through such networks [24].

In this paper we estimate the effect of distance on tie formation in a complete OSN. Next, we demonstrate the significance of spatial aspects of modularity of these networks by appropriate weighting of the links characterizing the connections between geographic units.

Description of the distance dependence of ties based on so called gravity models goes back at least to the work of Zipf [25]. It is assumed that the tie formation probability or the amount of communication flow is described as the product of the sizes of communities and a decaying function of the distance between them. Such modeling has found support by recent studies using telephone data [18, 19, 4, 26]. Spatial dependence then often embodies in regionally bounded network modules, for example in telephone-call networks [19], which suggested to be a general phenomenon [26]. Interestingly, distance dependence cannot be seen in the intensity of the contacts [8, 18].

The current paper will first present the dataset and some descriptive statistics. In the next section the distance effect on tie formation in the Hungarian OSN is compared to the characteristics of the Belgian telephone-call network [19]. Our results indicate that the OSN ties are less constrained by geographical distance. The following section introduces an analysis of the town-to-town network that is utilizing two different approaches to aggregate interpersonal OSN ties. This analysis reveals that the town-town tie weights are highly correlated by geographical distance. Moreover the networks have a modularity structure that follows different levels of administrative regions in the country, and the line of the river Danube.

## 2. Data and Methods

The iWiW (International Who Is Who) was launched on the 14th of April, 2002 and shortly became the most widely used OSN in Hungary and even the most visited national website at its peak in 2006. The website quickly gained significant scientific interest, for an early study see [27]. Later, the website could not meet the challenges of competing with market-leading OSNs (namely Facebook) and, after a long declining period, the service was shut down on June 30, 2014. In January 2013 there were over 4 million registered profiles, which covered around 40% of the total population of Hungary (10 million) and roughly two-third of the online population (aged 14 or more) at that time. This means that the vast majority of the online population has come into contact with iWiW over the decade of its life-cycle. A non-disclosure agreement signed by our research group and the data owner provides us with a unique access to the anonymized version of public profile data of all users, including basic demographic characteristics, a complete set of the connections within the OSN together with the date of their establishment, and the time of the user's last login. Unfortunately, the data owner did not give the permission to either extend the non-disclosure agreement or publish the source of the data but gave us the permission to publish an aggregated version of the data, on which our empirical exercise is based on.

Data also included self-reported information on the name of towns where users resided and the schools they attended. Although self-reports have been considered to be problematic [28], when localizing OSN users or social media content, it was compulsory to choose a location from a scroll-down menu upon registering a profile on iWiW. The residence could have been easily changed afterwards (as for example if users moved from one town to another), however there was no eligibility check or IP address-based control of this information. Thus, one might consider our location indicator based on user profiles as a biased and occasionally updated census-type data. The Supporting Information (S1.Figure and S2.Figure) contains an overview of settlement and regional structure of Hungary to make the interpretation of the results easier.

For our analysis in this paper we eliminated those 524,425 user profiles in which location was set to towns outside Hungary (for simplicity, we use the expression 'town' for all the settlements, including cities and villages). The top five countries that follow Hungary in terms of the number of profiles are Romania (167,198), Great Britain (55,461), the United States of America (35,966), Germany (34,732), and Serbia (19,941). The majority of the profiles located abroad (457,702) have at least one connection in Hungary. We also dropped those 193 users that had more than 10,000 connections because they might use the website for marketing purposes. This arbitrary threshold was set in order to distinguish those users from the data who have far more connections than can actually exist. Although these few profiles might not disturb the spatial pattern of the network, we decided to drop them because our argument focuses on social ties and not on the social media aspect of OSNs. .

We aggregate the friendship ties to a town-level. This data are available (http://datadryad.org/review?doi=doi:10.5061/dryad.33ps4) and the validity of a gravity-type modeling can be also verified [18-19]. The aggregation to a town-level description requires the introduction of weights (proportional to the number of individual links between towns). This weighted network will then be used to explore the spatial modularity of the OSN. Initial tie weights in the town-level network are the aggregated number of friendship between town $i$ and town $j$ (Table 1).

**Table 1. Aggregation of the iWiW network to a town-level.**

|  | USER-level network | TOWN-level network |
|---|---:|---:|
| Number of NODES | 4,078,513 | 2,558 |
| Number of TIES | 336,963,425 | 1,363,032 |
| Number of intra-town TIES | 186,237,393 | 2,558 |
| Number of inter-town TIES | 150,726,032 | 1,360,474 |

Data for the user-level network take into account all iWiW links. Data for the town-level network are aggregated: whenever there is at least one user-level link between two persons in different towns there is a link between those towns. As there are always intra-town links, the number of loops in the town network equals with the population of towns.

The total OSN population (4,078,513 users) is located in 2,558 towns; whereas there are 1,363,032 town–town connections. There are 336,963,552 user–user ties in total, out of which 186,237,520 ties remain within town borders and 150,726,032 ties are established between users from two distinct towns. The density of both networks were calculated using the formula $2 \times (\text{\# links}) / n(n-1)$, where $n$ is the number of nodes. For town-level density we get 0.42, for user-level density: 0.00405. The town-level network is very dense: 42% of all possible town-town ties exist; while the user-level network is naturally sparse: the density is smaller by two orders of magnitude. The town-level weighted degree distribution can be described by a power-law (Figure 1).

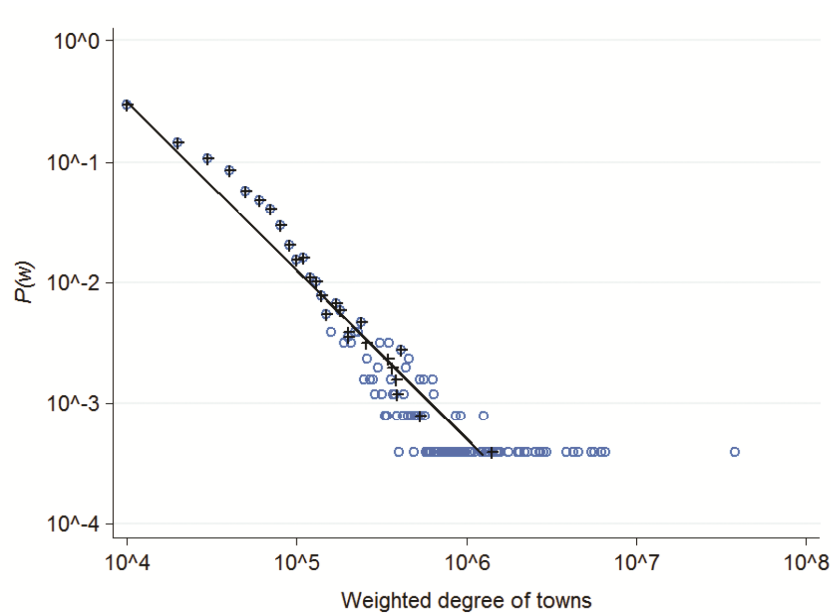

**Figure 1. Degree distribution.**
Weighted degree distribution of town network, loops excluded. Weights are the total number of user-to-user links between two towns. Weighted degrees were binned into $10^4$ intervals for $P(w)$ calculation; blue hollow circle symbols represent the bins; black plus symbols represent the mean of weighted degree by each unique value of $P(w)$. The slope of the solid line is -1.4, which fits $P(w)$ values with $R^2=0.66$.

# 3. Results

## 3.1 The distance effect

We first investigate the distance dependence of the link formation.

In order to address the distance decay effect on the intensity of OSN ties, we formulate a gravity model following the method of Lambiotte et al. [19] applied for telephone-call networks. Accordingly, we define $L(d)$ as the number of observed ties between users separated from each other by distance $d$; and $N(d)$ the number of possible ties at distance $d$. Then, we can calculate the probability that individuals have links to others given distance $d$ by the formula $P(d) = L(d)/N(d)$. Because it is not possible to measure distance between users residing in the same town in our data, only inter-town links were taken into consideration. In order to compare results to previous research [19], a 5 km resolution was used for binning distance distribution. We set an upper limit of $d$ at 480 km in the illustration of the results in order to avoid confusing readers with details originating from the Hungarian town structure [28].

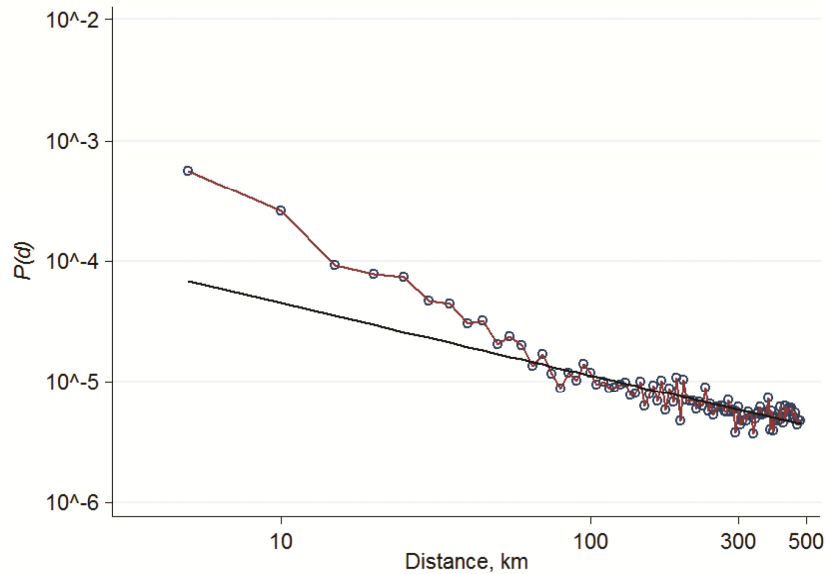

**Figure 2. The probability of links as function of distance.**
Probability $P(d)$ is plotted as function of distance on a log-log scale. The straight line depicts a power-law $d^{-0.6}$.

Comparing our results in Figure 2 with the Belgian telephone-call network, a few interesting observations can be made. The probability that users at small distance $d$ are linked is a magnitude higher in the OSN than in the telephone-call network. For example, for distance $d = 10$ km $P(d)$ equals to $10^{-3.5}$ in the OSN and was around $10^{-5}$ in the phone-call network [19]. Furthermore, we find a much slower decay with the distance in the OSN than it was previously reported for the phone-call networks. A power-law with $d^{-2}$ was repeatedly found for telephone-call networks [18, 19]. The exponent of distance decay in our data is -0.6, which means that geographical distance has a weaker effect on the OSN ties than on the telephone-calls. The Supporting Information (S3.Figure) illustrates tie probability as the function of distance, in which profiles located outside of Hungary are involved as well. The S3.Figure further supports our finding regarding the weak distance decay in OSNs.

## 3.2 Town-level weights

Spatial dependence is analyzed by utilizing two alternative edge weights. First, there is the natural weight in the town-level network:

$$w_{ij} = \# \text{ ties between towns } i \text{ and } j.$$

Here and in the following we disregard loops, i.e. $i \neq j$ always. We take the 10 base logarithm to handle the large variations in the weights:

$$w_{ij}^{(1)} = \log w_{ij}$$

This weight disregards the number of users in the specific settlements; therefore it gives an extra weight to highly populated towns. In order to take this effect into account we introduce a second weight:

$$w_{ij}^{(2)} = \log(w_{ij} / \overline{w}_{ij}),$$

where

$$\overline{w}_{ij} = \frac{s_i s_j}{\sum_{i,j}^{n} w_{ij}}.$$

Here $s_i = \sum_j^n w_{ij}$ is the strength of node $i$ and $\overline{w}_{ij}$ is the expected number of links between towns $i$ and $j$ based purely on the total number of links at those towns assuming random tie formation. We disregard loops in node strength $s_i$, $s_j$ and $\sum_j^n w_{ij}$ calculation. Note that $w_{ij}^{(2)}$ can be negative or positive depending on the ratio of the measured weight and the expected one.

The aim with $w_{ij}^{(2)}$ is to compare the strength of the connections between towns to a null model, which takes already into account the strengths of the towns [29]. The logic of this approach is similar to studies in sociology that filter out group size bias as social network determinants [31-33].

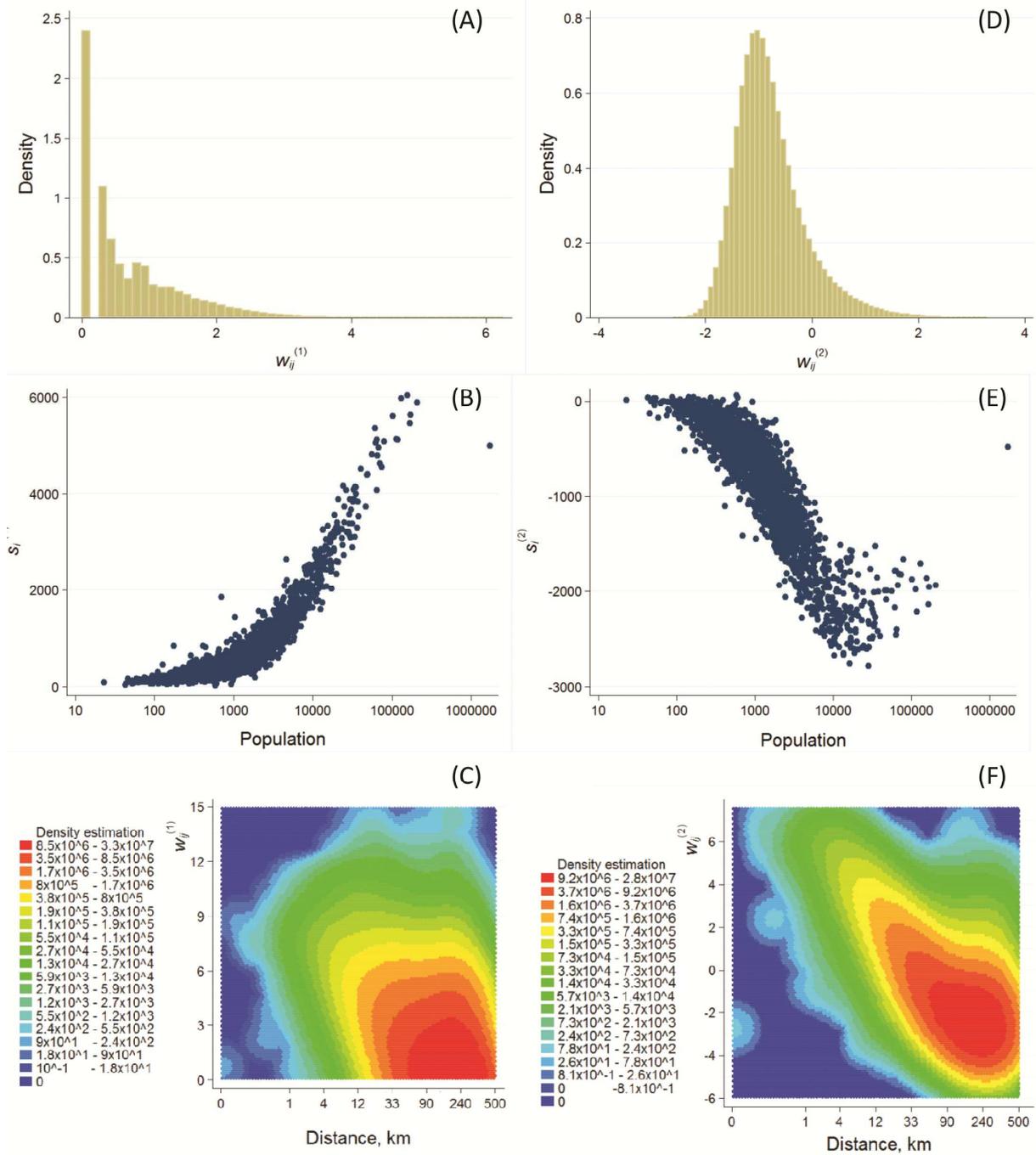

**Figure 3. Edge weight and node strength distributions.**
(A) The distribution of $w_{ij}^{(1)}$ weights is unimodal and with a maximum at the value of 1. The tail ends at 7.68 indicating that a large fraction of the ties represent very low number of connections compared to a small set of high links indicating large number of personal ties. (B) Node strength $s^{(1)}$ rises as population increases. The capital, Budapest with its 2 million inhabitants is an outlier. (C) The heat map of the density of $w_{ij}^{(1)}$ as a function of $w_{ij}^{(1)}$ and the distance between towns $i$ and $j$ shows a complex distribution that is dominated by a large number of weak ties between distant locations. (D) $w_{ij}^{(2)}$ has a unimodal distribution with values between –2.61 and 3.29 and with a modus at –0.77. (E) Node strength $s^{(2)}$ decreases as population increases but fluctuation is high across large towns. Budapest is again an outlier. (F) The heat map of the density of $w_{ij}^{(2)}$ as a function of $w_{ij}^{(2)}$ and the distance between towns *i* and *j* illustrates that the highest edge weights are between towns that are in geographical proximity.

Edge weight distributions in Figure 3A and Figure 3D provide distinct approaches to the OSN. The distribution of $w_{ij}^{(1)}$ is much skewed. There are very few strong ties in the $w_{ij}^{(1)}$ network compared to the large number of weak ties. On the other hand, the $w_{ij}^{(2)}$ distribution is two-tailed and resembles a Gaussian distribution; the mode is in the negative range.

The strength of node *i* in Figure 3B and 3E is defined as $s_i^{(1,2)} = \sum_j w_{ij}^{(1,2)}$. These two demonstrate again that the two edge weights capture different characteristics of the network. Node strength calculated from $w_{ij}^{(1)}$ increases as population grows suggesting a hierarchical topology, while node strength from $w_{ij}^{(2)}$ decreases as population increases. Interestingly, Budapest neither has extreme high strength with $w_{ij}^{(1)}$ nor extreme low strength with $w_{ij}^{(2)}$, although it stands out in terms of population.

Two-dimensional heat map of the density was calculated on a 100×100 grid as a function of edge weight and geographical distance with 10 grid bin size. Figure 3C exhibits a complex relation between distance and $w_{ij}^{(1)}$; a large number of weak ties between distant locations characterizes the two-dimensional distribution. However, the strongest ties are also found between distant locations. A much clearer distance-dependence emerges in the case of $w_{ij}^{(2)}$ in Figure 3F, where the highest positive edge values are among towns that are in close geographical proximity. One can also observe a growing variation of inter-town tie weights; maximum and minimum values decrease as distance grows until a certain threshold (distance ~ 33 km) after which positive weights depend only loosely on distance. This observation might uncover that distance decay on OSN ties is stronger within cohesive territories or commuting zones [34] than in large areas. However, those intercity ties that are extremely weak compared to the expected value ($w_{ij}^{(2)}$ below zero) seem to occur between distant places and these dominate the distribution.

## 3.3 Spatial modularity

Figure 4 shows the geography of the strongest links with the two different weights. The strongest edges in the $w_{ij}^{(1)}$ network are between large towns (Figure 4A). The majority of users are in these locations; consequently, the network is shaped along the settlement hierarchy, in which Budapest is the absolute center of the network and regional university towns also function as hubs. The spatial distribution of strongest edges in the network with $w_{ij}^{(2)}$ weights is entirely different (Figure 4C). The large weights are not anymore organized around Budapest and smaller regional centers also lose their prevalence as spatial hubs, while edges between small towns become strong. The inspection of the strongest ties implies that large towns tend not to establish edges that are above the expectation. Consequently, the densely grouped strong edges can be found in areas with fragmented settlement structure (west and northeast of the country), while the density of strong edges is low in areas where relatively larger settlements are located (southeast of the county) (see Supporting Information).

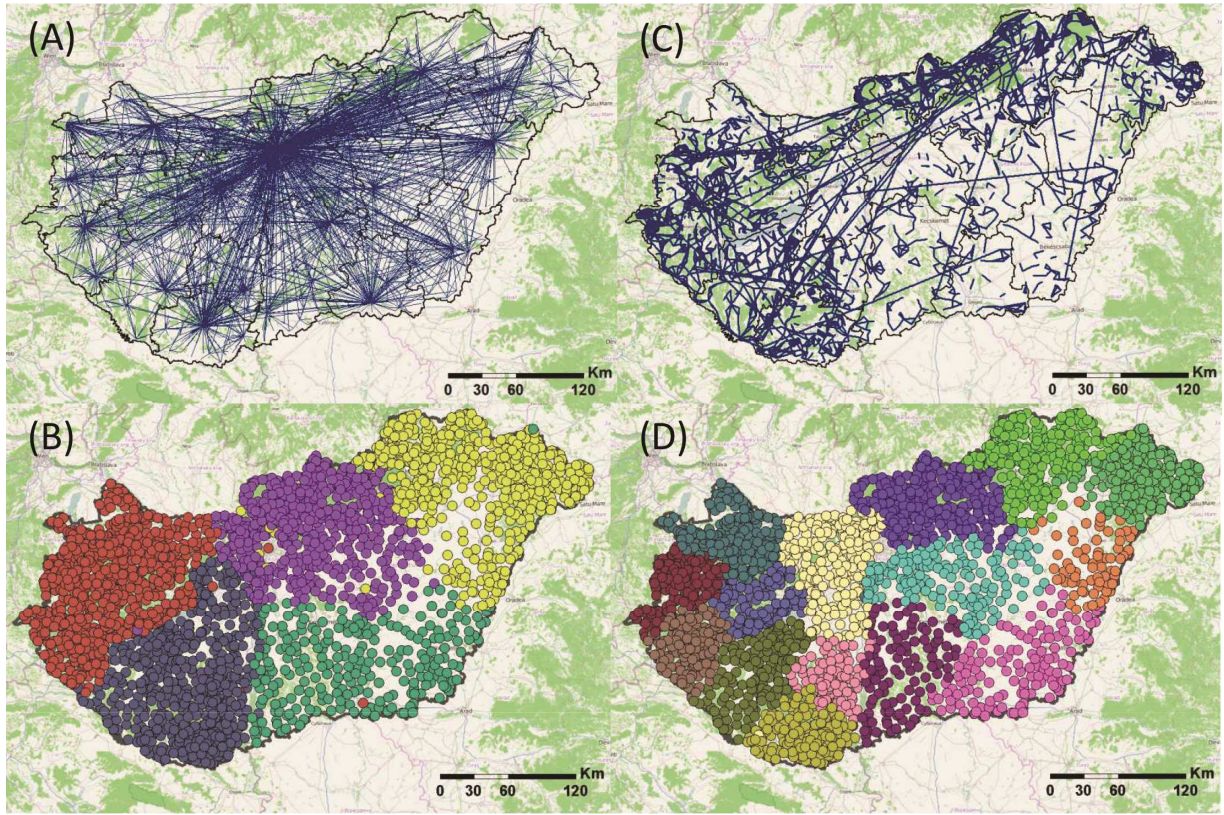

**Figure 4. Spatial structure and modularity.**
(A) The strongest 0.3% of all edges are depicted in $w_{ij}^{(1)}$ network (4,081 edges). (B) The Louvain method finds 5 modules in the $w_{ij}^{(1)}$ network; towns belonging to the same module are depicted with same colors. (C) The strongest 0.3% of all edges are depicted in $w_{ij}^{(2)}$ network (4,081 edges). (D) The illustrated community structure contains 14 modules in the $w_{ij}^{(2)}$ network. This community structure has the highest modularity index out of the Louvain algorithm runs we present in Table 2. Towns belonging to the same module are depicted with same colors.
Created by own data, with base map of OpenStreetMap cartography licensed as CC BY-SA.

Although the town network is very dense and towns have weak edges to distant towns with high probability, geographically-identifiable clustering might be present as well [26]. This may result in spatial modularity of the network, and if so, the network might fall into relatively cohesive spatial units. In order to identify clusters in the town network, the Louvain community identification method was used with resolution parameter set to 1 in both $w_{ij}^{(1)}$ and $w_{ij}^{(2)}$ networks [35]. The algorithm was run five times; we set 100 random restarts in each run and selected one community structure with the highest modularity index from each run. The maximum number of iterations in each restart was 20, the maximum number of levels in each iteration was 20, and the maximum number of repetition in each level was 50. Then we calculated the pair-wise Cramer's V index of the selected community structures, which takes its maximum at 1 when the compared module structures are identical.

**Table 2. Network modularity.**

| | Communities | Modularity | Towns in modules | | | Cramer's V | | | |
|---|---|---|---|---|---|---|---|---|---|
| | | | Min | Max | Mean | $w_{ij}^{(2)}$–1 | $w_{ij}^{(2)}$–2 | $w_{ij}^{(2)}$–3 | $w_{ij}^{(2)}$–4 |
| $w_{ij}^{(1)}$ | 5 | 0.209 | 254 | 720 | 511.6 | | | | |
| $w_{ij}^{(2)}$–1 | 14 | 0.373 | 72 | 285 | 182.7 | 1 | | | |
| $w_{ij}^{(2)}$–2 | 15 | 0.372 | 72 | 285 | 170.5 | 0.983 | 1 | | |
| $w_{ij}^{(2)}$–3 | 13 | 0.372 | 106 | 285 | 196.7 | 0.968 | 0.982 | 1 | |
| $w_{ij}^{(2)}$–4 | 14 | 0.371 | 72 | 285 | 182.7 | 0.954 | 0.965 | 0.953 | 1 |
| $w_{ij}^{(2)}$–5 | 15 | 0.372 | 72 | 270 | 170.5 | 0.984 | 0.960 | 0.981 | 0.963 |

Few large communities are identified in the $w_{ij}^{(1)}$ network compared to the $w_{ij}^{(2)}$ network. The five runs of the Louvain method finds exactly the same community structure in the $w_{ij}^{(1)}$ network, therefore we report the result of only one run. The community finding algorithm produced distinct community structures in the $w_{ij}^{(2)}$ network; therefore we report all five runs (the number after the hyphen denotes the sequence of the run). The pair-wise Cramer's V index in $w_{ij}^{(2)}$ network is always above 0.95.

Figures 4B and 4D illustrate that the community structure of the graphs are spatially based in both cases. Modularity is weaker (0.209) in the $w_{ij}^{(1)}$ network, in which five cohesive clusters are identified (Table 2). These clusters are very stable because exactly the same structure has repeatedly emerged in each of our five runs. Three large towns (Budapest, Szeged and Székesfehérvár) do not belong to their surroundings in terms of modularity but to the North-Western module depicted in red. The sizes of clusters are comparable to the NUTS level 2 regions (NUTS abbreviated from Nomenclature of Territorial Units for Statistics) or planning and statistical regions of the country (see S2.Figure). These were established by administrative reforms closely related to the statistical basis of European Union's development policy in 1999, although the Parliament refused to delegate significant functions to these regions in 2006. Our finding provides new evidence of the relevance of this regional scale in terms of social relations.

The $w_{ij}^{(2)}$ network shows a higher modularity index (0.373) and the Louvain algorithm results in 14 clusters. The community structures revealed by the five runs are not totally identical, but the values of the Cramer's V index are high, denoting considerable similarity. The somewhat weaker stability of module structure in the $w_{ij}^{(2)}$ network is the result of few communities breaking up and merging into neighboring other communities (see S4.Figure). Clusters in the $w_{ij}^{(2)}$ network reflect NUTS level 3 regions or counties of the country with few exceptions (see S2.Figure). These administrative regions were formulated in 1950 and functioned as important fields of economic and social planning for decades. The region around Budapest breaks into parts and these merge into neighboring regions. Interestingly, this separation occurs along the river Danube and the river also serves as a boundary for other communities in each of the five runs of the Louvain algorithm, which is another example supporting the idea that geographical barriers influence social networks [36].

# Discussion

In this paper we analyzed the complete iWiW network and found that, in spite of the cost neutrality of internet based connections, people establish distant online connections with lesser probability than proximate ones. However, geographical distance has smaller deflating power on the frequency of online friendship in our data than it was demonstrated previously using mobile call networks.

The fact that a large number of edges between small and distant towns exists may unveil further spatial characteristics that differentiate OSNs from other communication platforms. We argue that not only the cost of establishing a tie (as Borgatti et al [1] put it), but also costs for maintaining the tie are relevant; the latter clearly differs in telephone-call networks and OSNs. Tie establishment is a probably less selective process in OSNs than in telephone-call networks, both of which have been frequently analyzed in previous geography-related research [4, 13, 18-21, 23-24]. Our results indicate that online tools are also used for contacting persons, who are usually not reached by other media, while telephone contacts are mainly used as a communication channel for relationships, which are fostered by other means too. Users usually collect old friends and weak connections in OSNs and these links enhance the long distance weights. However, once a tie is established it is almost costless to maintain in an OSN; on the contrary, this is very costly in a phone-call interaction. Unfortunately, we do not have data about the frequency of contacts, which would be needed to prove the hypothesis related to the above reasoning: the long distance contacts are the less close relationships in the OSN too.

Another important observation of our study is that the OSN network is modular and the cohesive modules are based on geographical areas that coincide with administrative regions. However, different spatial modules emerge according to the alternative tie-weights used in the town-town network and reflect two levels of regional scale. Our results support previous arguments claiming that OSNs depend strongly on physical geography.

# Acknowledgments

The authors thank Norbert Barics and Zhongyuan Ruan for their help with the database. Helpful comments were received during seminar lectures organized by RECENS Group of Hungarian Academy of Sciences, International Business School Budapest, and Institute of Economics of Hungarian Academy of Sciences.

**Supporting Information files**

**S1.Figure.** Settlements and population structure in Hungary.

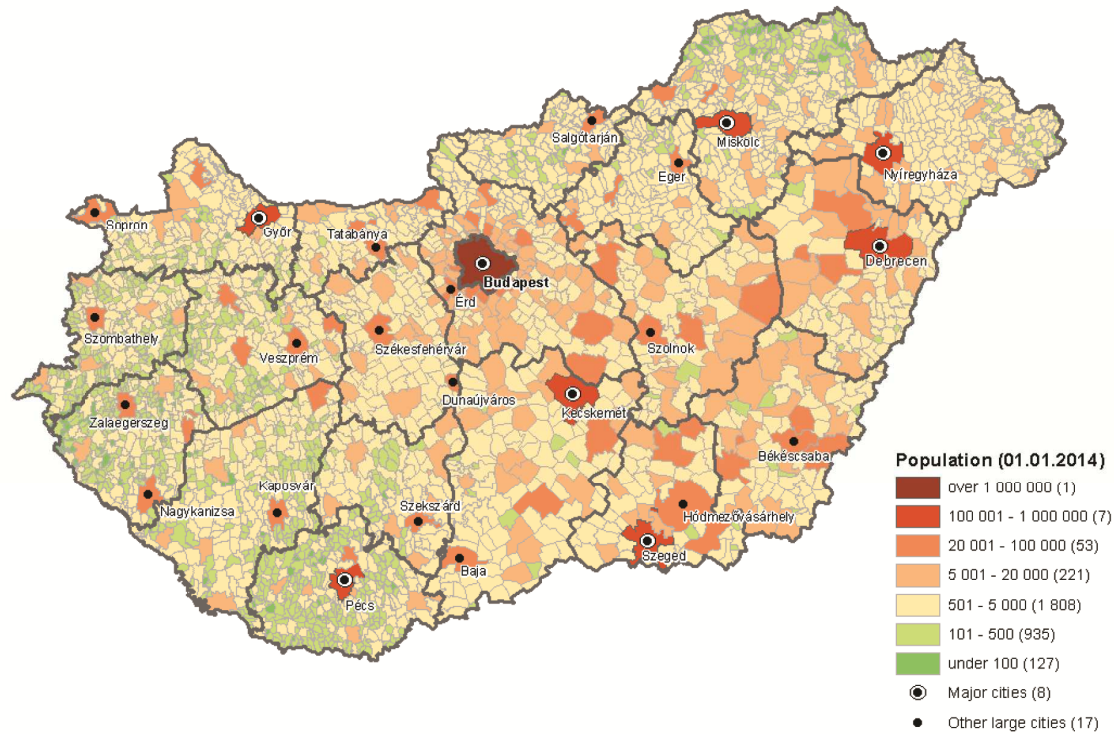

**Supporting Information 1: Settlements and population structure in Hungary.** Almost 1/5 of the country's population lives in Budapest. Hungary has 8 cities with population over 100.000 (including Budapest with 1.7 million people), 53 middle sized towns (with population between 20.001 and 100.000), 221 small-middle sized towns (with population between 5001 and 20.000), and also 2870 small towns and villages with less than 5000 inhabitants. There are remarkable regional differences in town size: many small towns can be found in the Western part of the country and in the North-East. The average town size is significantly larger in the South-East. The largest cities (over 100.000 people) are unevenly located, only two of them can be found in Transdanubia (Western part of the country).
Own creation of the authors, based on census data of the Hungarian Central Statistical Office.

**S2.Figure.** Regions and counties in Hungary.

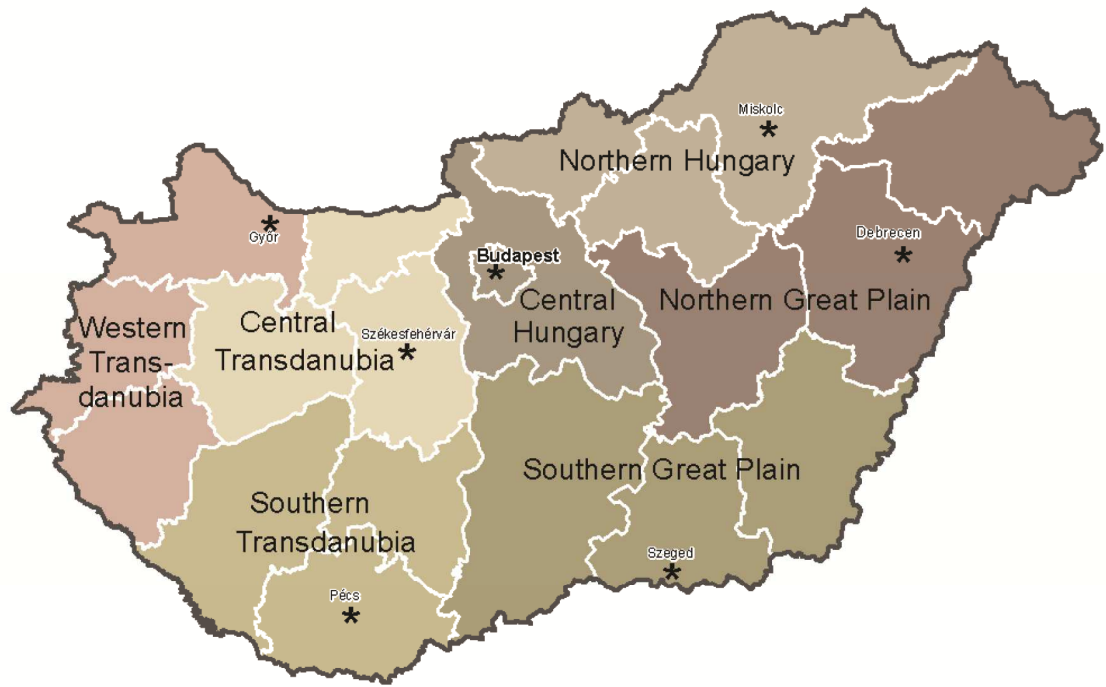

**Supporting Information 2: Regions and counties in Hungary**.
The Nomenclature of territorial units for statistics, abbreviated as NUTS is a geographical nomenclature subdividing the territory of the European Union into regions at three different levels (NUTS 1, 2 and 3, respectively, moving from larger to smaller territorial units). Above NUTS 1 is the 'national' level of the Member State. The 7 NUTS2 regions are depicted by color and their names and centers are presented in the map. 19 NUTS3 regions – so called counties – are separated by white borders in the map.
Own creation of the authors based on http://ec.europa.eu/eurostat/statistics-explained/index.php/Glossary:Nomenclature_of_territorial_units_for_statistics_%28NUTS%29.

**S3.Figure**. The probability of links as function of distance, involving all profiles regardless of country location.

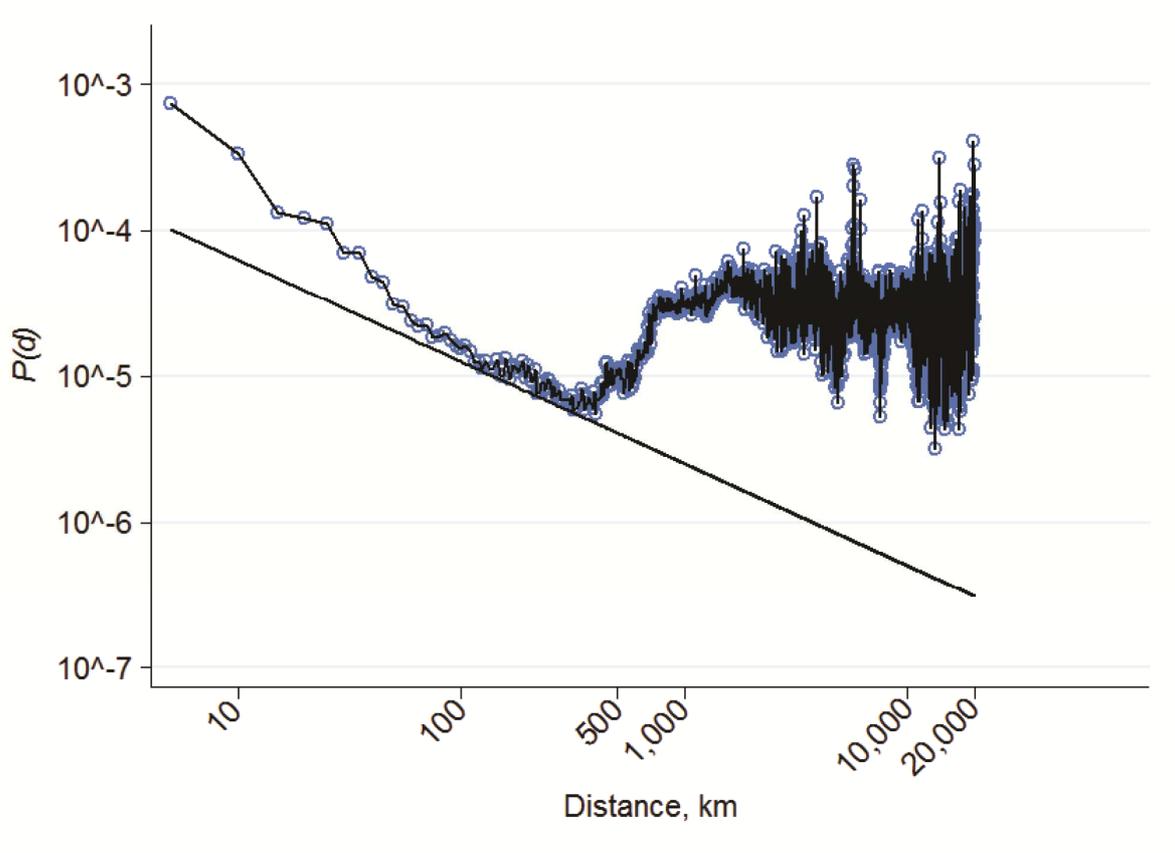

**Supporting Information 3: The probability of links as function of distance, involving all profiles regardless country location.**

Same as Figure 2 in the main text but including the locations outside of the country. Clearly, the approximately power law dependence of $P(d)$ on $d$ breaks down at the frontier of Hungary ($d>500$) as there the organization of users follow different principles. These are partly Hungarians living in neighbouring countries (Romania and Serbia) or, to large extent, persons, who left the country recently and have their contacts according to their original locations (UK, US and Germany).

**S4.Figure.** Community structures in the $w_{ij}^{(2)}$ network found by five separate runs of the Louvain algorithm.

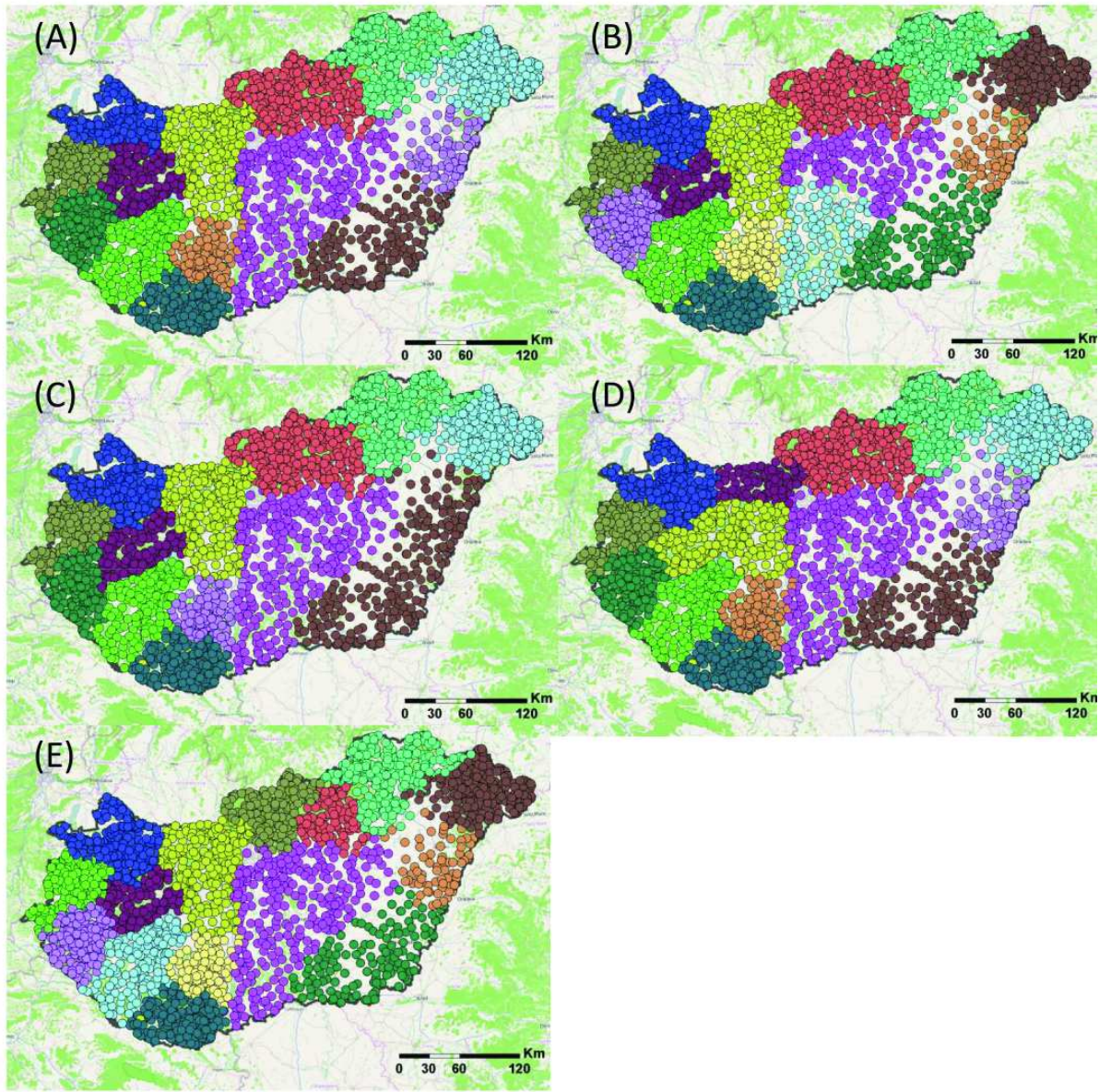

**Supporting Information 4: Community structures in the network found by five separate runs of the Louvain algorithm.** Pest county does not constitute a separate community in any of the runs. (A) The community structure $w_{ij}^{(2)}$-1 depicts 14 modules. Four pairs of counties are merged into one community: Bács-Kiskun and Jász-Nagykun-Szolnok; Békés and Csongrád; Fejér and Komárom-Esztergom; Heves and Nógrád. (B) The community structure $w_{ij}^{(2)}$-2 depicts 15 modules. Three pairs of counties are merged into one community: Békés and Csongrád; Fejér and Komárom-Esztergom; Heves and Nógrád. (C) The community structure $w_{ij}^{(2)}$-3 depicts 13 modules. Békés, Csongrád and Hajdú-Bihar are merged into one community and three pairs of counties are merged into one community: Bács-Kiskun and Jász-Nagykun-Szolnok ; Fejér and Komárom-Esztergom; Heves and Nógrád. (D) The community structure $w_{ij}^{(2)}$-4 depicts 14 modules. Four pairs of counties are merged into one community: Bács-Kiskun and Jász-Nagykun-Szolnok; Békés and Csongrád; Fejér and Veszprém; Heves and Nógrád. (E) The community structure $w_{ij}^{(2)}$-5 depicts 15 modules. Three pairs of counties are merged into one community: Bács-Kiskun and Jász-Nagykun-Szolnok; Békés and Csongrád; Fejér and Komárom-Esztergom.
Created by own data, with base map of OpenStreetMap cartography licensed as CC BY-SA.